\documentclass[aps,prb,twocolumn,showpacs,superscriptaddress]{revtex4}
\usepackage{epsfig,amsmath,amssymb}
\usepackage{graphicx}
\usepackage[english]{babel}

\begin{document}

\title{A tunable coupler with ScS quantum point contact  \\
to mediate strong interaction between flux qubits   }

\author{A.A. Soroka} 
\affiliation{National Science Center ``Kharkov Institute of Physics
and Technology'', Akhiezer Institute for Theoretical Physics, 61108
Kharkov, Ukraine}
\author{V.I. Shnyrkov}
\affiliation{B. Verkin Institute for Low Temperature Physics and
Engineering, National Academy of Sciences of Ukraine,
61103 Kharkov, Ukraine}%

\begin{abstract}
In this paper we propose a kind of quantum inductance couplers
(QUINC) which represents a superconducting loop closed by ScS
quantum point contact, operating in deep quantum low-temperature
regime to provide tunable (Ising-type) ZZ interaction between flux
qubits. This coupler is shown to be well tunable by an external
control magnetic flux and to provide large inter-qubit interaction
energies $\!|J/k_{\rm B}|\!\simeq1$\,K thus being very promising as
a qubit-coupling device in a quantum register as well as for
studying fundamental low-temperature quantum phenomena. %
Some entanglement measures of a two-qubit system are analyzed as
functions of inter-qubit interaction strength.
\end{abstract}
\maketitle

\section{Introduction}\vspace{-1mm}

Handling interaction between basic elements (qubits) of a quantum
computer is one of key problems for implementation of computation
algorithms therein \cite{Steane,Kilin,Valiev}. %
In a computer register including $N$ qubits, 
computation operations are generally elements of the group $U(2^N)$
of unitary transforms of a superpositional state vector in the
$2^{N}$-dimensional tensor-product Hilbert state space of all qubits
with the basis:
$\{s_1\}\!\otimes\!\{s_2\}\!\otimes\ldots\otimes\!\{s_N\}$,
$\{s_i\}\!=\!\left\{\left|\uparrow\right\rangle_i;\left|\downarrow\right\rangle_i\right\}$
being basis sets of separate qubits. Unitary transforms of a quantum
register state vector in $2^{n}$-dimensional subspaces %
(forming the unitary group $U(2^n)$) %
which are produced by turning on interaction within certain
subsystem of $n$ qubits during some timespan and which realize
specific computation operations are called $n$-qubit gates. A
universal set of gates for quantum computation is such that
generates all possible unitary transformations in the full Hilbert
vector space and thus suffices for implementation of an arbitrary
algorithm. In quantum informatics, the Brylinski's theorem states
that a universal set of computation gates may be constituted from
all one-qubit gates (providing local unitary transformations of
separate qubits) and any non-primitive, that is entangling,
two-qubit gate \cite{Brylinski}. The most known entangling two-qubit
gate in quantum informatics is the CNOT gate \cite{Barenco}.
Different entangling two-qubit gates are interrelated involving
one-qubit gates \cite{KL}.

So, a physical system used for building a quantum register should
provide a tuning of interaction energy between pairs of qubits to
form regulated entangled two-qubit states, with the possibility of
turning on/off inter-qubit interaction \cite{DiVincenzo}. A coupler
with a quantum point contact (QPC) we propose realizes ZZ two-qubit
tunable interaction generating a class of entangling gates that, as
follows from the aforesaid, produce a universal set for quantum
computation. Furthermore, control of interaction between the quantum
coherent systems is of great interest to study fundamental effects
of entanglement and correlations in the states of coupled qubits and
qutrits \cite{Steane}.

Lately, a system of two coupled Josephson, and particularly, flux
qubits has been intensively studied \cite{Clarke}. In first papers
simple systems of flux qubits were studied, with qubits located
side-by-side and coupled by constant coefficients of mutual
inductances $M_{ij}$, determined by geometry of the relative
position and self-inductances of qubits \cite{Izmalkov,Majer}. At
the same time, new versions of coupling elements were proposed
theoretically \cite{Plourde,Brink,Wilhelm} which, based on dc- and
rf- SQUIDs,  enabled to vary the magnitude and the sign of magnetic
interaction of flux qubits to be coupled by means of an external
control parameter (the bias current in dc SQUID and the external
magnetic flux applied to the loop of rf SQUID). These sign- and
magnitude-tunable qubit-coupling elements were implemented in the
Refs.\,\onlinecite{Hime,Harris} for the first time, but obtained
magnitudes of qubits' interaction energies, proportional to
respective dynamical Josephson inductances, were rather small (about
50\,mK), being insufficient for coupling qubits with high tunnel
energy splitting \cite{Shnyrkov1}.

In this paper we propose a tunable coupler to provide strong ZZ
interaction between flux qubits using the quantum inductance of a
superconducting loop closed by ScS quantum point contact
\cite{Shnyrkov2}. A quantum inductance coupler (QUINC) is a
nonlinear quantum system being in some of its eigenstates, such that
additional unwanted entanglement between this state and states of
the coupled qubits could be excluded. On this account, the coupler
working state, with largely varying quantum inductance (curvature of
the energy vs.\!\! flux dependence), is found that forms in a
three-well asymmetric potential of the quantum loop in such a way as
to be localized in a side well of the potential and tunnel to its
central well very weakly. We study the quantum inductance of such
coupler, which determines interaction energy between two flux
qubits, depending on the external magnetic flux through the
coupler's loop as the control parameter. Also we analyze some
widely-used entanglement measures for a system of two identical
qubits with tunable coupling energy of ZZ type as functions of its
magnitude.

\section{Model and numerical analysis}

\begin{figure}[t!]
\centering %
\includegraphics[width = 1.0 \columnwidth ]{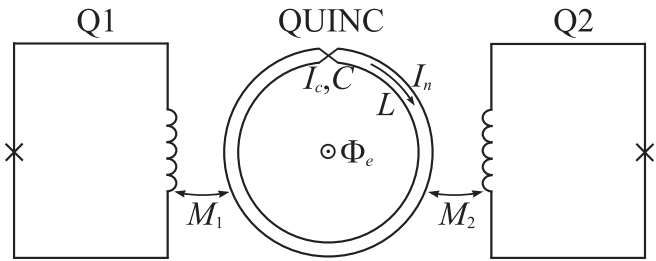} %
\caption{\label{fig01}\vspace{-0mm} %
The QUINC with ScS quantum point contact which mediates coupling
between the qubits Q1 and Q2. The loop of the coupler has inductance
$L$ and the contact of the coupler has self-capacitance $C$ and
critical current $I_{c}$.
}%
\end{figure}

Let's consider a system of two flux qubits having tunnel splittings
$\Delta\!E_{01}^{(1)}\!=\!2\Delta_1$ and
$\Delta\!E_{01}^{(2)}\!=\!2\Delta_2$ in points of symmetry of their
two-well potentials, which are inductively coupled to the coupler
via transformers of magnetic flux with coefficients of mutual
inductance $M_1$ and $M_2$, respectively (Fig.\,\ref{fig01}). The
coupler we select is a superconducting Nb 3D-loop closed by a clean
ScS quantum (atomic-size) point contact, such that
$d\!\sim\!\lambda_{F},d\!\ll\!l,\xi_0$, where $d$ is the contact
dimension, $\lambda_{F}$ is the electron wave length, $l$ is the
electron elastic mean free path, $\xi_0$ is the superconducting
coherence length \cite{Agrait,Beenakker}. Such 3D-loop of the
coupler may be topologically the same as that we used for designing
the qutrit \cite{Shnyrkov2}, but with a different set of parameters:
the loop inductance $L$, the self-capacitance $C$ and the critical
current $I_{c}$ of the ScS contact ($L$ and $I_c$ determining a
value of the nonlinearity parameter $\beta_L\!=\!2\pi LI_c/\Phi_0$,
$\Phi_0=h/2e$ is the flux quantum), that is necessary to establish
peculiar properties of the coupler's energy structure. Under
operation conditions, the coupler is initialized by means of
external flux $\Phi_e$ to certain energy level $E_{n}(\Phi_e)$,
which nonlinear dependence on control parameter $\Phi_e$ determines
coupling strength between qubits.

\vspace{-0mm}

It was shown in Ref.\,\onlinecite{Brink}, that energy of Ising-type
ZZ interaction of two flux qubits, mediated by a coupler as a
nonlinear quantum element, is
\begin{equation} \label{EQ01}
\begin{array}{c}
\displaystyle{ %
J(\Phi_{e})=-\chi(\Phi_{e}) M_1\!M_2 I_1\!I_2, } \vspace{2mm}\\%
\displaystyle{ %
\chi(\Phi_{e})\!\equiv\!\frac{1}{L_Q}\!=\!-\frac{dI_n}{d\Phi_{e}}\!=\!\frac{d^2 E_{n} }{d\Phi_{e}^2}\!=\!%
\frac{1}{L}\!\left(\!1\!-\!\frac{d\Phi}{d\Phi_{e} } \right)\!, %
} %
\end{array}
\end{equation}
where $I_1,I_2$ are the superconducting currents of the basis states
$\left\{\left|\uparrow\right\rangle_i;\left|\downarrow\right\rangle_i\right\}$
($i\!=\!1,2$) of the first and the second qubits; $E_n(\Phi_{e})$
and $I_n(\Phi_{e})$ are the energy of the coupler operation level
and circulating in its loop superconducting current as functions of
magnetic flux $\Phi_{e}$ applied to the loop, respectively;
$\Phi(\Phi_{e})\!=\!\Phi_{e}\!-\!LI_{n}(\Phi)$ is the
internal magnetic flux in the loop; %
$\chi(\Phi_{e})\!=\!E_n''(\Phi_e)$ is the local curvature of the
coupler operation level called the susceptibility or, equivalently,
the reciprocal quantum inductance $L_Q^{-1}$ of the level. %
As seen from Eq.\,(\ref{EQ01}), interaction energy $J(\Phi_e)$
between two flux qubits is determined by the susceptibility function
$\chi(\Phi_{e})$, to within constant coefficients of mutual
inductances and qubit currents. So we calculate the $\chi(\Phi_{e})$
function of the QUINC with ScS atomic-size contact to analyze its
specific coupling properties. Note that positive values of the
curvature ($\chi\!>\!0$) correspond to ferromagnetic coupling of
qubits (FM, $J\!<\!0$), while negative
values ($\chi\!<\!0$) to antiferromagnetic coupling (AFM, $\!J>\!0$). %

It should be emphasized that a quantum inductance coupler, being a
quantum element, must function in one-dimensional Hilbert space, so
as not to get entangled with the qubits, thus resulting in effective
four-dimensional Hilbert state space of the qubit subsystem. It can
be achieved provided that (i) the operation level $E_{n}(\Phi_{e})$
is weakly-superpositional, (ii) the distances between $E_{n}$ and
the neighboring levels $E_{n-1},E_{n+1}$ substantially exceed the
splittings $\Delta\!E_{01}^{(1)},\Delta\!E_{01}^{(2)}$ of the
qubits. Note that quantum coherent adiabatic regime of the coupler
functioning allows to eliminate decohering influence of
quasiparticle currents, inherent to a classical system (SQUID), on
the qubit dynamics.

To analyze the QUINC with ScS atomic-size contact (such that the
parameter $g\!=\!E_{J}/E_{C}\!=\!\Phi_0 I_{c} C/(2\pi e^2)\!\gg\!
1$) at the bath temperature $T$ considerably less than any distances
between adjacent energy levels of the quantum system (thus
validating a zero-temperature approximation), we use the
flux-representation Hamiltonian in the form \cite{Shnyrkov1,Leggett} 
\begin{equation} \label{EQ02}
\begin{array}{c}
\displaystyle{ %
\hat{H}_{C} =\frac{\hat{P}^2}{2M} + \hat{U}(f;f_e)=
} \vspace{2mm}\\%
\displaystyle{ %
=-\frac{\hbar^2}{2M} \frac{\partial ^2}{\partial f^2}
+ \frac{\Phi_0 I_c}{2\pi}\! %
\left[ -2 \left|\cos(\pi f) \right|+ \frac{2\pi^2 (f-f_{e})^2}{\beta_{L}} \right], %
} 
\end{array}
\end{equation} %
where $f= \Phi/\Phi_0$ and $f_e= \Phi_e/\Phi_0$ are the normalized
internal magnetic flux $\Phi$ in the coupler loop and external
magnetic flux $\Phi_e$ applied to the loop, $M=\Phi_0^2 C$ is the
respective effective mass. The quantum dynamical observable of the
internal magnetic flux in the coupler loop is given by an operator
$\hat{\Phi}$ conjugated to an operator
$\hat{Q}$ of charge in the contact capacitance: %
$[\hat{\Phi },\hat{Q}]=i\hbar$ \cite{Leggett}. The key feature of
Hamiltonian (\ref{EQ02}) is its singular potential $U(f;f_e)$
following from the non-sine current-phase relation of ScS quantum
point contact \cite{Beenakker}:
$$ I_s(\varphi) = I_c\sin(\varphi /2)\,\textrm{sgn}[\cos(\varphi /2)],\,\,%
I_c=\frac{\pi\Delta_0}{eR_N}= N\frac{e\Delta_0}{\hbar },
$$
where $\Delta_0$ is the superconducting energy gap, $R_N$ is the
normal-state resistance of the contact, $N$ is an integer %
($\varphi$-dependence is identical with that of the classical ScS
contact \cite{Kulik}). The critical current of ScS atomic-size
contact is quantized in consequence of quantization of the contact
conductance $R_N^{-1}$ in units of $G_0\!=\!2e^2/h$, that was
observed experimentally \cite{Agrait}. Note that such an effective
quantum Hamiltonian (\ref{EQ02}) of the superconducting loop closed
by the ScS contact, with its Josephson energy
($U_J(\varphi)\!=\!-(I_c\Phi_0/\pi)|\cos(\varphi /2)|$) having the
singular peculiarity and the dissipation vanishing at zero
temperature, satisfactorily describes the experiments both (i) on
macroscopic quantum tunneling phenomena in the loop with the clean
ScS contact \cite{Shnyrkov3,Shnyrkov4} and (ii) on coherent quantum
superposition of macroscopically distinct states in the three-well
potential of the superconducting qutrit \cite{Shnyrkov2}. %
It is important to note with regard to the inner quasiparticle
mechanism of dissipation in superconductors with Josephson contacts
at zero temperature, that quantum fluctuations in both the ScS and
SIS contacts with characteristic frequencies $\omega\ll \Delta_0$
were shown \cite{Khlus,Eckern} to correspond to equilibrium
fluctuations of a quantum oscillator without damping and with a
renormalized contact capacitance. In the opposite limit of high
frequencies $\omega\gg \Delta_0$ the impact of quasiparticle
excitations in the contacts gives rise to an effective linear
dissipation described by the Caldeira-Leggett model. Due to large
magnitude of the Nb gap ($\Delta_0/h = 330\,$GHz), the quasiparticle
dissipation can be neglected in the considered coupler.

\begin{figure}[t!]
\centering %
\vspace{-7mm} %
\includegraphics[width = 1.0\columnwidth ]{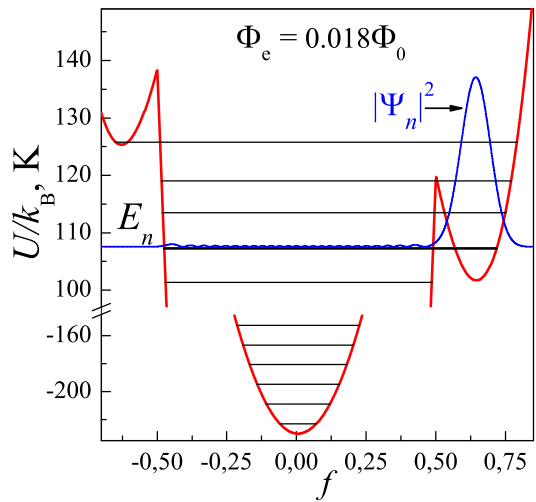} %
\vspace{-14mm} %
\caption{\label{fig02} %
The coupler potential $U(f)/k_{\rm B}$ (in units of temperature)
at  $\beta_L=4.4$, $C=3.1\,$fF, $f_e=0.018$; %
$E_n$ is the operation level of the coupler, %
$|\Psi_n|^2$ is the squared wave function of the respective quantum
state.
}%
\end{figure}

\begin{figure}[t!]
\centering %
\vspace{-5mm} %
\includegraphics[width = 1.0\columnwidth]{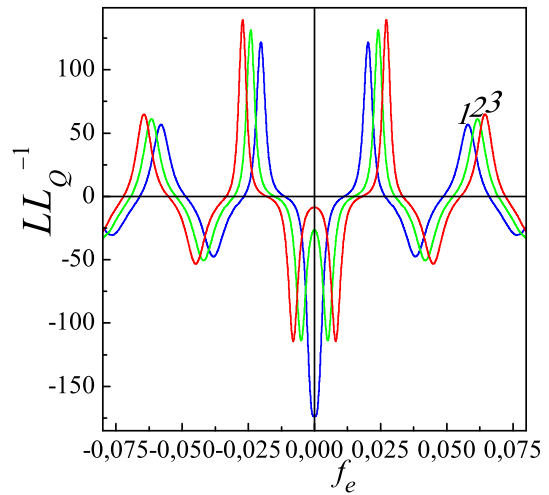} %
\vspace{-12mm} %
\caption{\label{fig03} %
The function $(LL_Q^{-1})(f_e)$ for $L\!=\!0.3$\,nH, %
$\beta_L\!=\!4.4$, and the values of $C$: %
\textit{1} -- 3.0\,fF, \textit{2} -- 3.05\,fF, \textit{3} -- 3.10\,fF. %
}%
\end{figure}

The solutions of the stationary Schr\"{o}dinger equation
\begin{equation} \label{EQ03}
  \hat{H}_{C}(f;f_{e})\,\Psi(f)=E(f_{e})\,\Psi(f)
\end{equation}
with Hamiltonian (\ref{EQ02}) describe wave functions $\Psi(f)$ and
energies $E(f_e)$ of the stationary states of the coupler at a
specified external magnetic flux $f_e$. %
Let us consider solutions of Eq.\,(\ref{EQ03}) with the following
coupler parameters: $L\!=\!0.3\,$nH, $C\!=\!3.1\,$fF,
$\beta_L\!=\!4.4$ ($I_c\!\approx\!4.8\,$$\mu A$) and
$f_e\!=\!0.018$, corresponding to asymmetric three-well potential
$U(f;f_e)$, shown in Fig.\,\ref{fig02}. In this configuration, we
are interested in the state with minimal energy level $E_n$ lying
both in the right-side and the central wells of the potential.
Experimentally, this state $\Psi_n(f)$ may be obtained from the
initial ground state in the three-well symmetric potential at
$\Phi_e\!=\!\Phi_0$ as a result of decreasing $\Phi_e$ to the
required value. With decreasing $\Phi_e$, a central potential well
transforms into a side well, where the quasiclassical state with a
large quantum number ($n\!=\!25$ for $E_n$ in Fig.\,\ref{fig02}) is
localized. The level $E_n$ lies far below the potential barrier top,
$(U_b -E_n)/k_{\rm B}\!\approx\!12.7\,$K, and separated from
neighboring energy levels by a considerable interval $\Delta\!E$: %
$(E_n\!-\!E_{n-1} )/k_{\rm B}\!\approx\!(E_{n+1}\!-\!E_n)/k_{\rm
B}\!\approx\! 6\,$K. As the state $\Psi_n(f)$ is almost entirely
localized in the right well, it proves to be utterly
weakly-superpositional. For the parameters in Fig.\,\ref{fig02}
[such that $\chi(f_e)\!\approx\!0$] the probability of the state
$\Psi_n(f)$ being in the right well
$P_R\!=\!\int_R\!|\Psi_n(f)|^2df\!\approx\!0.97$, and in the central
well $P_C\!=\!\int_C\!|\Psi_n(f)|^2 df\!\approx\! 0.03$. The
quantity $P_C$ characterizes a superpositional leakage of the state
$\Psi_n(f)$ to the central well. Further the coupler is
characterized in an operation $f_e$-range which is defined by a
criterion of ensuring a desired fixed range of its normed
susceptibility [$(LL_Q^{-1}\!)\!(f_e)\!\in\!(-10\,.\,.\,10)$, see
below]. When varying $f_e$ away from the point $f_{0}$ of zeroing of
the $\chi$-function (at a given $\beta_L$), the $P_C$ magnitude
increases; for curves in Fig.\,\ref{fig03}, the maximal
$P_C\!=\!0.04$ in the operation $f_e$-range. With increase in
$\beta_L$ the potential barrier between the right and the central
wells grows (and so grows the depth of the right well), that causes
reducing of the $P_C$; e.g. for $\beta_L=4.6$ the maximal
$P_C\!=\!0.02$ in the operation range. At that the $P_R$ approaches
nearer to unity.

It should be pointed out that under actual conditions the coupler
state will be always to some extent decohered (being in the
so-called partially-coherent state), because of finite nonzero
environment temperature and influence of unavoidable decohereing
(noise-originating) factors. We assume the situation in which the
influence of small superpositional probability $P_C$ of the coupler
being in the central well on its entanglement with the qubits is
smoothed away due to the finite noise-induced phase dispersion of
the wave function $\Psi_n(f)$. Thus, weak superpositionality of the
coupler operation state and its large separation from the
neighboring states (multiply exceeding the tunnel splittings of
qubits) can make the coupler a well-defined entity not entangled
with the qubits.

Because of small setting times of the operation state during
variation of the external magnetic flux, estimated as $\tau\!\sim\!
h/\!\Delta\!E\!\sim\!10^{-11}\,$s, the coupler will be described by
the adiabatic susceptibility $\chi(\Phi_e)$ even at rather high switching rates. %
Similarly to stability of the base superposition state of the qutrit
toward relaxation to underlying states \cite{Shnyrkov2}, the coupler
operation level will be the same stable under certain conditions. It
is possible due to designing of the coupler loop in the form of a
high-quality three-dimensional toroidal superconducting cavity with
no resonant modes at frequencies corresponding to that of
transitions from the operation
level to underlying ones. %

\begin{figure}[t!]
\hspace{-1.1cm} %
\vspace{-0mm} %
\includegraphics[width = 1.0\columnwidth]{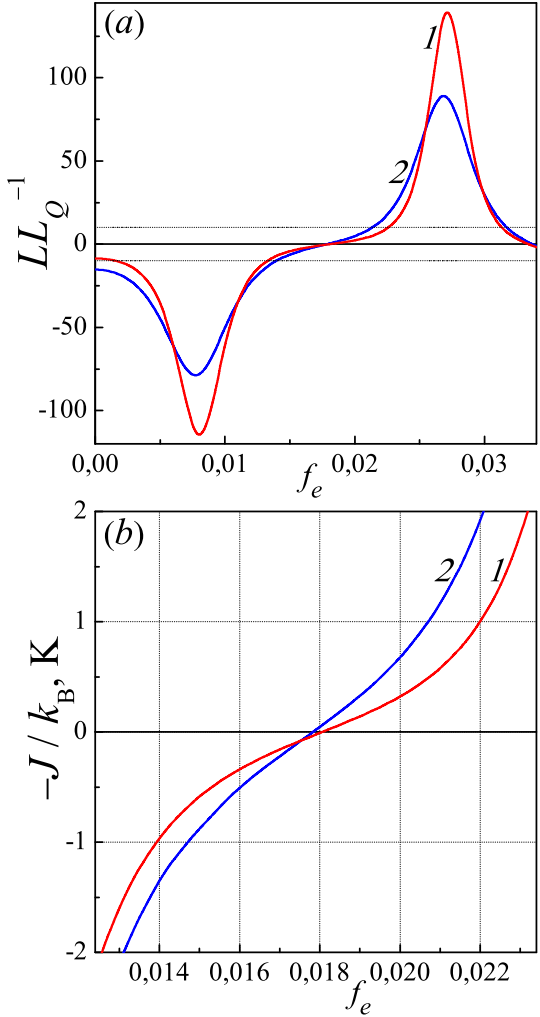} %
\caption{\label{fig04} %
(a) The function $(LL_Q^{-1})(f_e)$ for $L\!=\!0.3$\,nH, %
$C\!=\!3.10$\,fF and the values of $\beta_L$: \textit{1} -- 4.4,
\textit{2} -- 4.2; %
(b) The function $-J(f_e)/k_{\rm B}$ for $L\!=\!0.3$\,nH, $C\!=\!3.10$\,fF, %
$M_1\!=\!M_2\!=\!0.03\,$nH, $I_1\!=\!I_2\!=\!0.8\,\mu$A, %
and the values of $\beta_L$: \textit{1} -- 4.4, \textit{2} -- 4.2. %
}%
\end{figure}

For practical analysis, the function $(LL_Q^{-1}\!)\!(f_e)$ of the
susceptibility (i.e. $L_Q^{-1}$) normed to the loop inductance $L$
is convenient \cite{Shnyrkov2}. It is obtained by numerical solving
Eq.\,(\ref{EQ03}), finding $E_n(f_{e})$ and its double
differentiation by the
$f_e$ parameter: $(LL_Q^{-1}\!)\!(f_e)\!=\!(L/\Phi_0^2)d^2E_n\!/\!df_e^2$. %
In Fig.\,\ref{fig03} the function $(LL_Q^{-1}\!)\!(f_e)$ is shown
for $\beta_L\!=\!4.4$ and several values of the ScS-contact
capacitance. With increasing $C$, the curve $(LL_Q^{-1}\!)\!(f_e)$
shifts right parallel to the x-axis. %
At that a form of the curve does not change qualitatively in the
operation $f_e$-range of the coupler. In Fig.\,\ref{fig04}(a) the
function $(LL_Q^{-1}\!)\!(f_e)$ is shown
for two different values of $\beta_L$. %
As discussed above, with increasing $\beta_L$ the state $\Psi_n(f)$
still more localizes in the right well. Nonlinearity of the
$E_n(f_e)$ function increases, and so the $(LL_Q^{-1}\!)\!(f_e)$
rises in absolute value, with its slope at the x-axis decreased.
This decrease of the susceptibility slope about the $f_0$ point can
be compensated by widening of the coupler operation range as
necessary. Fig.\,\ref{fig04}(b) shows the function $-J(f_e)/k_{\rm
B}$ (see Eq.\,(\ref{EQ01})) of the mediated by the coupler
inter-qubit interaction energy (in temperature units and with minus
sign for convenience) for the same set of the coupler parameters as
in Fig.\,\ref{fig04}(a) and the characteristic absolute values of
coupler-qubit mutual inductances $M_1\!=\!M_2\!=\!0.1L$ and of the
basis currents in qubits. 
It is seen that at chosen parameters the variation of the control
external flux in the relatively small range
$f_e\!\in\!(0.014\,.\,.\,0.022)$ ensures the wide range of variation
$(-J/k_{\rm B})\!\in\!(-1\,.\,.\,1$\,K)  of the inter-qubit coupling
energy.

\section{Entanglement measures} %

Let us now analyze properties of the two-qubit system with ZZ
interaction $J(f_e)$ mediated by the considered coupler (see
Fig.\,\ref{fig01}). Having obtained eigenstates and density matrix
of the system, we will explicitly find entanglement measures, widely
used for characterization of the entangled states
\cite{Wootters,SM,Bennett}, and consider an example of the
entanglement time variation in the studied system.

The effective Hamiltonian of the system in the tensor-product basis
\begin{equation} \label{EQ04}
\begin{array}{c} \hspace{-0.0mm}
\displaystyle{ %
 \left\{ \left| \uparrow \right\rangle_1 ; \left| \downarrow \right\rangle_1 \right \} \!\otimes\! %
 \left\{ \left| \uparrow \right\rangle_2 ; \left| \downarrow \right\rangle_2 \right \} = %
 \left\{ \left| \uparrow\uparrow \right\rangle ; \left|\uparrow\downarrow\right\rangle ; %
 \left|\downarrow\uparrow\right\rangle ;
 \left|\downarrow\downarrow\right\rangle \right\}
}%
\end{array}
\end{equation}
(the numbers of arrows correspond to omitted qubit indices in
"two-arrow"\, states) of the four-dimensional state space, close to
qubits' degeneracy points, has the form
\begin{equation} \label{EQ05}
\begin{array}{c}
\displaystyle{ %
\hat{H}_{12} = (-\varepsilon_1\hat{\sigma}_z\!-\!\Delta_1\hat{\sigma}_x)\otimes \hat{\sigma}_0 + %
               \hat{\sigma}_0 \otimes (-\varepsilon_2 \hat{\sigma}_z\!-\!\Delta_2\hat{\sigma}_x) + %
} \vspace{1.5mm}\\
\displaystyle{ + J(f_e)\,\hat{\sigma}_z\!\otimes\hat{\sigma}_z = } \vspace{2mm}\\ %
\displaystyle{ =\!\! \left(\!
                   \begin{array}{cccc}
                     -\varepsilon_1\!-\!\varepsilon_2\!+\!J & -\!\Delta_2 & -\!\Delta_1 & 0 \\
                     -\!\Delta_2 & -\varepsilon_1\!+\!\varepsilon_2\!-\!J & 0 & -\!\Delta_1 \\
                     -\!\Delta_1 & 0 & \varepsilon_1\!-\!\varepsilon_2\!-\!J & -\!\Delta_2 \\
                     0 & -\!\Delta_1 & -\!\Delta_2 & \varepsilon_1\!+\!\varepsilon_2\!+\!J \\
                   \end{array}\!
                 \right)\!.
}
\end{array}
\end{equation}
Here $\varepsilon_1$, $\varepsilon_2$ are the energy biases of first
and second qubits relative to their degenerate energy levels in
symmetric two-well qubit potentials, $\hat{\sigma}_x,
\hat{\sigma}_z, \hat{\sigma}_0$ are the Pauli matrices. The
fourth-order characteristic equation for finding eigenvalues of
$\hat{H}_{12}$ [Eq.(\ref{EQ05})]
\begin{equation} \label{EQ06}
\begin{array}{c}
\displaystyle{ %
E^4 -
2(\varepsilon_1^2\!+\!\varepsilon_2^2\!+\!\Delta_1^2\!+\!\Delta_2^2\!+\!J^2)E^2-(8\varepsilon_1\varepsilon_2J)E }\vspace{1mm}\\ %
\displaystyle{ %
+[\varepsilon_1^4\!+\!\varepsilon_2^4\!+\!J^4\!+\!(\Delta_1^2\!-\!\Delta_2^2)^2\!+\!2J^2\!(\Delta_1^2\!+\!\Delta_2^2)+ }\vspace{1mm}\\%
\displaystyle{ %
+2(\Delta_1^2\!-\!\Delta_2^2)(\varepsilon_1^2\!-\!\varepsilon_2^2)\!-\!2(\varepsilon_1^2\varepsilon_2^2\!+\!\varepsilon_1^2J^2\!+\!\varepsilon_2^2J^2)]=0 %
}%
\end{array}
\end{equation}
can be solved in radicals, e.g. by Ferrari's method. The most
interesting with relation to the quantum mechanics is the symmetric
case of unbiased qubits ($\varepsilon_1\!=\!\varepsilon_2\!=\!0$)
with maximal superposition and entanglement effects. In this case
Eq.\,(\ref{EQ06}) is simplified to a biquadratic equation and the
two-qubit system has the eigenenergies
\begin{equation} \label{EQ07}
\begin{array}{c}
\displaystyle{ E_i= \mp\sqrt{J^2+(\Delta_1\pm\Delta_2)^2},\quad i=1,2,3,4. } %
\end{array}
\end{equation}
We focus on the still more symmetric and indicative case of unbiased
qubits with equal tunnel amplitudes
($\Delta_1\!=\!\Delta_2\!=\!\Delta\!=\!\Delta\!E_{01}/2$), %
in which the eigenvalues of $\hat{H}_{12}$ are
\begin{equation} \label{EQ08}
\begin{array}{c}
\displaystyle{ E_1\!=\! -\sqrt{J^2+4\Delta^2 } },\, %
\displaystyle{ E_2\!=\! -J},\,%
\displaystyle{ E_3\!=\! J},\, %
\displaystyle{ E_4\!=\! \sqrt{J^2+4\Delta^2} }, %
\end{array}
\end{equation}
and the corresponding normed eigenstates are
\begin{equation} \label{EQ09}
\begin{array}{c}
\displaystyle{ \left|\Psi_1\right\rangle = a\left|\uparrow\uparrow\right\rangle + %
                                           b\left|\uparrow\downarrow\right\rangle + %
                                           b\left|\downarrow\uparrow\right\rangle +
                                           a\left|\downarrow\downarrow\right\rangle },\, 2(a^2+b^2)\!=\!1,\vspace{1mm}\\  %
a\!=\!\frac{\Delta}{\sqrt{J^2\!+\!4\Delta^2}}\left(\! 1+\frac{J}{\sqrt{J^2+4\Delta^2}}\right)^{\!-\frac{1}{2}}\!, \, %
b\!=\!\frac{1}{2}\left(\!1+\frac{J}{\sqrt{J^2+4\Delta^2}}\right)^{\!\frac{1}{2}}; \vspace{3mm}\\ %
\displaystyle{ \left|\Psi_2\right\rangle = \frac{1}{\sqrt{2}}\left( \left|\uparrow\downarrow\right\rangle - %
                                                                    \left|\downarrow\uparrow\right\rangle \right); \hspace{2mm}%
\left|\Psi_3\right\rangle = \frac{1}{\sqrt{2}}\left( \left|\uparrow\uparrow\right\rangle - %
                                                                    \left|\downarrow\downarrow\right\rangle \right); } \vspace{3mm}\\ %
\displaystyle{ \left|\Psi_4\right\rangle = a'\!\left|\uparrow\uparrow\right\rangle + %
                                           b'\!\left|\uparrow\downarrow\right\rangle + %
                                           b'\!\left|\downarrow\uparrow\right\rangle +
                                           a'\!\left|\downarrow\downarrow\right\rangle },\, 2(a'^2+b'^2)\!=\!1, \vspace{1mm}\\  %
a'\!=\!\frac{\Delta}{\sqrt{J^2\!+\!4\Delta^2}}\left(\!1-\frac{J}{\sqrt{J^2+4\Delta^2}}\right)^{\!-\frac{1}{2}}\!, %
b'\!=\!-\frac{1}{2}\left(\!1-\frac{J}{\sqrt{J^2+4\Delta^2}}\right)^{\!\frac{1}{2}}\!.

\end{array}
\end{equation}
The density operator of the system in pure (at zero temperature)
eigenstates $\left|\Psi_i\right\rangle$ (\textit{i}=1,2,3,4) has the
form
\begin{equation} \label{EQ10}
\begin{array}{c}
\displaystyle{ %
\hat{\rho}=|\Psi_i\rangle\langle\Psi_i| }\,.%
\end{array}
\end{equation}
For definiteness sake, we focus on the ground state $|\Psi_1\rangle$
of the system. Taking into account Eqs.\,(\ref{EQ09}), we obtain
the density matrix $\rho_{kl}= \langle k|\hat{\rho}|l\rangle$ %
(in the basis (\ref{EQ04})) in the ground state
\begin{equation} \label{EQ11}
\begin{array}{c}
\displaystyle{ %
\rho_{kl} = \begin{pmatrix}
              a^2 & ab & ab & a^2 \\
              ab & b^2 & b^2 & ab \\
              ab & b^2 & b^2 & ab \\
              a^2 & ab & ab & a^2 \\
            \end{pmatrix}
}
\end{array}
\end{equation}
(note that components of $\rho_{kl}$ are real as the initial
tunneling amplitudes $\Delta_1,\Delta_2$ were chosen real, without
loss of generality for considering stationary states and
entanglement measures). The density matrix of a pure state, such as
(\ref{EQ11}), has the only nonzero eigenvalue $\lambda=1$  (as
$\hat{\rho}^2=\hat{\rho}$ for it and thus $\lambda^2=\lambda$). %
The reduced density operators for either of the two qubits are
defined as partial traces of $\hat{\rho}$ over basis vectors of the
other qubit, that is %
\begin{equation} \label{EQ12}
\begin{array}{c}
\displaystyle{%
\hat{\rho}_1\!=\!{\rm Tr}_{(2)}\hat{\rho}= \left\langle k|\hat{\rho}|k \right\rangle\!,\,\,  %
|k\rangle = \left|\uparrow\right\rangle_2, \left|\downarrow\right\rangle_2, %
} \vspace{2mm}\\%
\displaystyle{%
\hat{\rho}_2\!=\!{\rm Tr}_{(1)}\hat{\rho}= \left\langle k|\hat{\rho}|k \right\rangle\!,\,\,  %
|k\rangle = \left|\uparrow\right\rangle_1, \left|\downarrow\right\rangle_1. } %
\end{array}
\end{equation}
So, the reduced density matrices of the first and the second qubits
in the system described by (\ref{EQ11}) are
\begin{equation} \label{EQ13}
\begin{array}{c}
\displaystyle{%
\rho_1\!=\!\rho_2\!=\!\begin{pmatrix}
                1/2 & 2ab \vspace{1.0mm}\\
                2ab & 1/2 \\
              \end{pmatrix}\!;\,\,
\lambda_1\!=\!2ab+\frac{1}{2},\lambda_2\!=\!-2ab+\frac{1}{2}, } %
\end{array}
\end{equation}
with $\lambda_1,\lambda_2$ being the eigenvalues of the equal
matrices.

Now we proceed to analysis of entanglement measures. A two-qubit
state
\begin{equation}\nonumber 
\begin{array}{c}
\displaystyle{ \left|\Psi\right\rangle = A\left|\uparrow\uparrow\right\rangle + %
                                         B\left|\uparrow\downarrow\right\rangle + %
                                         C\left|\downarrow\uparrow\right\rangle +
}                                        D\left|\downarrow\downarrow\right\rangle   %
\end{array}
\end{equation}
is entangled, that is by definition not decomposable as a tensor product of states of two qubits %
\begin{equation}\nonumber 
\begin{array}{c}
\displaystyle{%
(\alpha\left|\uparrow\right\rangle_1 + \beta\left|\downarrow\right\rangle_1)\otimes %
(\gamma\left|\uparrow\right\rangle_2 + \delta\left|\downarrow\right\rangle_2), %
}%
\end{array}
\end{equation}
if and only if the inequality $AD-BC\ne 0$ is fulfilled, which
follows straight from the definition (\ref{EQ04}) of the
two-qubit basis. Entanglement measure of two-qubit states %
\begin{equation} \label{EQ14}
\begin{array}{c}
\displaystyle{%
{\mathcal E}_0={\rm 4}|AD-BC|^{{\rm 2}},
}%
\end{array}
\end{equation}
is called the tangle, having such plain structure \cite{Wootters}. %
For the ground state  $\left|\Psi_1\right\rangle$, given by
(\ref{EQ09}), we have:
\begin{equation} \label{EQ14'}
\begin{array}{c}
\displaystyle{%
{\mathcal E}_0=4(a^2-b^2)^2 = \frac{J^2}{J^2+4\Delta^2}=\frac{J_0^2}{1+J_0^2}, } \tag{\ref{EQ14}$'$} %
\end{array}
\end{equation}
where $J_0\equiv J/(2\Delta)=J/\Delta\!E_{01}$ is the two-qubit
interaction energy $J$ normed to the tunnel splitting
$\Delta\!E_{01}$ of the qubits.

There is a widespread density-matrix formalism measure of
entanglemet \cite{SM}:
\begin{equation} \label{EQ15}
\begin{array}{c}
\displaystyle{%
{\mathcal E}_D=\frac{4}{3}{\rm Tr}\!\left\{ (\hat{\rho}-\hat{\rho}_1\!\otimes\!\hat{\rho}_2)^2\right\} }. %
\end{array}
\end{equation}
It quantifies the Hilbert-Schmidt distance between a density matrix
and a tensor-product matrix from reduced density matrices of the two
qubits. Substituting matrices (\ref{EQ11}), (\ref{EQ13}) into
(\ref{EQ15}) we get for the state $\left|\Psi_1\right\rangle$:
\begin{equation} \label{EQ15'}
\begin{array}{c}
\displaystyle{%
{\mathcal E}_D\!=\!1\!-\!\frac{32}{3}b^2\!+\!\frac{128}{3}b^4\!-\!\frac{256}{3}b^6\!+\!\frac{256}{3}b^8\!=\!\frac{{\mathcal E}_0(2+{\mathcal E}_0)}{3}.%
}  \tag{\ref{EQ15}$'$} %
\end{array}
\end{equation}
The relation between ${\mathcal E}_D$ and ${\mathcal E}_0$ in
(\ref{EQ15'}) is strict for the pure states \cite{KSB}.

One more popular measure is information-theoretic one called the
entanglement of formation, defined as \cite{Bennett}:
\begin{equation} \label{EQ16}
\begin{array}{c}
\displaystyle{%
{\mathcal E}_S\!=\!S(\hat{\rho}_1\!)=\!S(\hat{\rho}_2)\!=\!-{\rm
Tr}\!\left\{
\hat{\rho}_1{\rm log}_2\hat{\rho}_1  \right\}\!=\!-\sum_{i=1}^2\lambda_i{\rm log}_2\lambda_i. }%
\end{array}
\end{equation}%
With $\lambda_1,\lambda_2$ from (\ref{EQ13}), Eq.(\ref{EQ16})
gives ${\mathcal E}_S$ for the state $\left|\Psi_1\right\rangle$. %
The measure ${\mathcal E}_S$ is related to the tangle ${\mathcal
E}_0$ for the pure states as well \cite{Wootters}. %
As follows from (\ref{EQ16}), the pure states themselves have
${\mathcal E}_S\!=\!0$, seeing that their density matrices have one
nonzero eigenvalue $\lambda\!=\!1$. On the other hand, the reduced
density matrices of two subsystems of a system in an entangle pure
state correspond to the mixed states. This property of efficient
states of the subsystems is treated as their mutual "measuring"\,
each other as parts of an entangled system.

\begin{figure}[t!]
\centering
\includegraphics[width = 1.0\columnwidth]{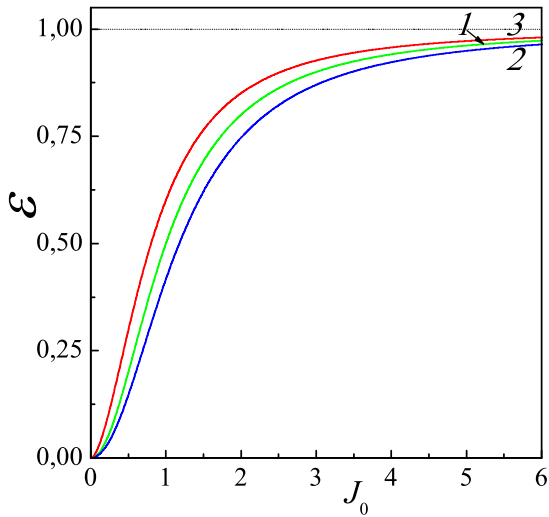} 
\caption{\label{fig05} The entanglement measures of the two-qubit
system in the ground state $\left|\Psi_1\right\rangle$
as functions of the dimensionless interaction parameter $J_0$: %
$\textit{1} - {\mathcal E}_0(J_0)$, $\textit{2} - {\mathcal
E}_D(J_0)$, $\textit{3} - {\mathcal E}_S(J_0)$. %
}%
\end{figure}

All the entanglement measures ${\mathcal E}_0, {\mathcal E}_D$ and
${\mathcal E}_S$ vanish in the tensor-product states, being positive
otherwise, and maximize to unity in the most entangled states (such
as states of the Bell's basis). In Fig.\,\ref{fig05}, the ${\mathcal
E}_0,{\mathcal E}_D,{\mathcal E}_S$ as functions of the qubits'
interaction parameter $J_0$, given by Eqs. (\ref{EQ14'}),
(\ref{EQ15'}), (\ref{EQ16}), are shown for the considered two-qubit
system in the ground state $\left|\Psi_1\right\rangle$. Due to
symmetry, exactly the same expressions hold for the highest excited
state $\left|\Psi_4\right\rangle$. At $J_0\!=\!0$
($a\!=\!b\!=\!1/2$, $a'\!=\!-b'\!=\!1/2$), i.e. at vanishing
interaction strength between the qubits, the states
$|\Psi_1\rangle$, $|\Psi_4\rangle$ decompose into the product of the
form
\begin{equation}\label{EQ17}
\begin{array}{c}
\left|\Psi_1\right\rangle =
\frac{1}{\sqrt{2}}(\left|\uparrow\right\rangle_1+\left|\downarrow\right\rangle_1)\!\otimes\!
\frac{1}{\sqrt{2}}(\left|\uparrow\right\rangle_2+\left|\downarrow\right\rangle_2), \vspace{1mm}\\
\left|\Psi_4\right\rangle =
\frac{1}{\sqrt{2}}(\left|\uparrow\right\rangle_1-\left|\downarrow\right\rangle_1)\!\otimes\!
\frac{1}{\sqrt{2}}(\left|\uparrow\right\rangle_2-\left|\downarrow\right\rangle_2),
\end{array}
\end{equation}
so that all the entanglement measures vanish, the qubits are
disentangled. Note, the reduced density matrices $\rho_1$ and
$\rho_2$ in Eq.(\ref{EQ13}) describe pure states in this case
(otherwise mixed ones). With increasing absolute value of $J_0$, all
the entanglement measures steadily increase. In the characteristic
point $J_0\!=\!1$: ${\mathcal E}_0(1)\!=\!0.5$, ${\mathcal
E}_D(1)\!=\!0.4$, ${\mathcal E}_S(1)\!=\!0.6$. At
$J_0\!\rightarrow\!\infty$, ${\mathcal
E}_0=1\!-\!J_0^{-2}\!+\!\mbox{o}(J_0^{-4})$.
And so, at increasing $|J|\!\gg\!\Delta$ (FM-coupling), %
the ground and the highest excited states tend to
\begin{equation}\label{EQ18}
\begin{array}{c}
\left|\Psi_1\right\rangle\!\rightarrow\!
\frac{1}{\sqrt{2}}(\left|\uparrow\uparrow\right\rangle\!+\!\left|\downarrow\downarrow\right\rangle),\,\,%
\left|\Psi_4\right\rangle\!\rightarrow\!
\frac{1}{\sqrt{2}}(\left|\uparrow\downarrow\right\rangle\!+\!\left|\downarrow\uparrow\right\rangle),
\end{array}
\end{equation}
the Bell's states having the maximal unity entanglement %
(while the eigenstates
$\left|\Psi_2\right\rangle\!,\!\left|\Psi_3\right\rangle$, %
being the Bell's states, have the unity entanglement at arbitrary
$J$).

The eigenstates $\left|\Psi_i\right\rangle$ (\ref{EQ09}) are
stationary, varying in time as $\exp(-iE_it/\hbar )$ with the
eigenenergies $E_i$ (\ref{EQ08}) (\textit{i}=1,2,3,4). So the
entanglement of the eigenstates is constant in time, as seen from
Eq.(\ref{EQ14}) for ${\mathcal E}_0$. Let's consider a simple
dynamical model in which the two-qubit interaction under study can
be turned on and off instantly. It simulates a situation, when
interaction turning on/off timespan is much less than the system's
characteristic dynamical time. Suppose, the interaction $J$ is
turned on between the qubits which are initially disentangled,
specifically, in the product state $\left|\Psi_1\right\rangle$ from
Eq.(\ref{EQ17}). Applying eigendecomposition of the Hamiltonian,
we get the time-varying state vector
\begin{equation}\label{EQ19}
\begin{array}{c}
\displaystyle{%
\left|\Psi\!(t)\right\rangle\!=\!e^{-i\hat{H}_{12}t/\hbar}\!\left|\Psi\!(0)\right\rangle\,, %
\langle k\left|\Psi\!(t)\right\rangle\!=\!(A,B,B,A),}\vspace{1mm}\\%
\displaystyle{ A\!=\!a(a+b)e^{-iE_1t/\hbar}+a'\!(a'+b')e^{iE_1t/\hbar}},\vspace{0mm}\\%
\displaystyle{B\!=\!b(a+b)e^{-iE_1t/\hbar}+b'\!(a'+b')e^{iE_1t/\hbar}\,,}
\end{array}
\end{equation}
and the tangle
\begin{equation}\label{EQ20}
\begin{array}{c}
\displaystyle{%
{\mathcal E}_0(t)\!=\!\frac{J_0^2}{1+J_0^2}\!\left[\!\frac{(\cos(2\pi t/t_0\!)-1)^2}{1+J_0^2} + \sin^2(2\pi t/t_0\!) \right]}, \vspace{0mm}\\%
\end{array}
\end{equation}
where $t_0\!=\!h/\!(2\sqrt{J^2\!+\!\Delta\!E_{01}^2})$ is its
oscillation period. At $J_0\!\le\!1$, time of transition from the
initial disentangled state to the maximally entangled one, with
${\mathcal E}_0^{max}\!=\![2/\!(J_0+J_0^{-1}\!)]^2,$ is
$t_\varepsilon\!=\!t_0/2$ (otherwise,
$t_\varepsilon=(t_0/\!2\pi)\arccos(-J_0^{-2}\!)\!>\!t_0/4$ and
${\mathcal E}_0^{max}\!=\!1$). So, from a dynamical viewpoint,
characteristic time of entangling of two initially disentangled
subsystems is determined by their interaction energy $J$:
$t_\varepsilon\!\sim\!\hbar/J$ at $J\!\sim\!\Delta$. And the mere
fact of entanglement between subsystems becomes possible because of
their interaction (${\mathcal E}_0(t)\!=\!0$ at $J\!=\!0$). Suppose
that the interaction $J_0\!=\!1$ is turned off at the moment
$t_\varepsilon$. Then the respective state vector %
$(0,\frac{i}{\sqrt{2}},\frac{i}{\sqrt{2}},0)$ %
will start to oscillate as %
$\langle k\left|\Psi(t)\right\rangle\!=\!(a_f,b_f,b_f,a_f)$, %
with $a_f\!=\!-\frac{\sin2\pi t/t_f}{\sqrt{2}}$, $b_f\!=\!\frac{i\cos2\pi t/t_f}{\sqrt{2}}$ %
and the period $t_f\!=\!h/\!\Delta\!E_{01}$, having constant %
${\mathcal E}_0(t)\!=\!1$, that illustrates the conception of
quantum nonlocality (though the latter suggesting that consistent
non-paradoxical description of interaction of quantum systems
requires the theoretical setting of relativistic quantum field
theory\cite{RQFT}).

\section{Summary}

We have analyzed the quantum inductance coupler based on a
superconducting loop with ScS quantum point contact, which is
intended to provide the tunable ZZ interaction between
superconducting flux qubits. The following features of this coupler
should be pointed out that favorably distinguish it from the
analogues by principle of operation. These are (i) the relatively
small operation range $\Delta f_e\!\sim\!0.01$ of the coupler
controlling flux, ensuring the wide range of strengths $J(f_e)$ and
hence promoting reduction in operational times of the coupler; (ii)
almost symmetric form of the function $J(f_e)$ relative to the point
of its vanishing in the operation range; (iii) large attainable
absolute values of the inter-qubit interaction strength
$|J(f_e)/k_{\rm B}|\!\sim 1\!$\,K. These features of the QUINC with
ScS QPC make it optimal for coupling the superconducting niobium
flux qubits with ScS QPCs as well having the tunnel splittings
$\Delta\!E_{01}/k_{\rm B}\!\sim\!1\!-\!2$\,K, or
$\Delta\!E_{01}/h\!\sim\!20\!-\!40$\,GHz \cite{Shnyrkov1} (notice
that properly designed superconducting Josephson circuits with QPCs
may serve as high-quality qubits, qutrits, couplers and quantum
detectors\cite{Shnyrkov2} thus having great potential for the
quantum information science). In such a way, having available flux
qubits with the splittings $\Delta\!E_{01}/k_{\rm
B}\!\simeq\!1\!$\,K and
couplers providing energies $|J/\!k_{\rm B}|\!\simeq\!1$\,K, %
a multi-qubit quantum system with tunable inter-qubit coupling
energies can be constructed, which will behave strongly coherently
at temperatures $T\!\sim\!10^{-2}$\,K,  enabling implementation of
quantum gates.

We showed behavior of the widely-used entanglement measures for the
eigenstates of the symmetric two-qubit system with ZZ inter-qubit
interaction as functions of the interaction strength $J$. With
different formal definitions, all the considered measures behave
qualitatively similarly in agreement with physical intuition, viz.
steadily increase with increasing $J$ in the system's ground and
highest excited states. Also time variation of the entanglement
(${\mathcal E}_0$-measure) in the analyzed two-qubit system was
illustrated within an elementary model of piecewise-constant $J(t)$
function. With increasing $J$, the entanglement time $t_\varepsilon$
reduces that favors to increase the rate of computational operations
in the system of coupled qubits. Simultaneously, the considered
instance shows that time-controlling of the entanglement is not
simple problem requiring modeling of the time dependence of
inter-qubit interaction strength and realizing it experimentally to
perform desired two-qubit gates.

\begin{acknowledgements}
We thank O.G. Turutanov for helpful discussions.
\end{acknowledgements}

\end{document}